\documentclass[a4paper,11pt]{article}
\usepackage{pos}
\usepackage{xcolor} 
\usepackage[multiple]{footmisc} 
\usepackage{booktabs} 
\usepackage{bm} 
\usepackage{multirow} 
\usepackage{siunitx}

\title{Expected performance of the ALTO particle detector array designed for 200 GeV - 50 TeV gamma-ray astronomy}
\ShortTitle{Expected performance of the ALTO particle detectors}

\author*[a]{Mohanraj Senniappan}
\author[a]{Yvonne Becherini}
\author[b,a]{Michael Punch}
\author[c]{Satyendra Thoudam}
\author[a]{Tomas Bylund}
\author[a]{Ga\v{s}per Kukec Mezek}
\author[d]{Jean-Pierre Ernenwein}

\affiliation[a]{Department of Physics and Electrical Engineering, Linnaeus University,\\
35195 Växjö, Sweden}

\affiliation[b]{Universit\'e de Paris, CNRS, Astroparticule et Cosmologie, \\
 F-75006 Paris, France}

\affiliation[c]{Department of Physics, Khalifa University,\\
PO Box 127788, Abu Dhabi, United Arab Emirates}

\affiliation[d]{Aix Marseille Univ, CNRS/IN2P3, CPPM,\\
Marseille, France}

\emailAdd{mohanraj.senniappan@lnu.se}

\abstract{The CoMET is an R\&D project aiming to design a very-high-energy (VHE) gamma-ray observatory sensitive to energies above $\sim$\SI{200}{\GeV}. The science goals include continuous observation of soft-spectrum VHE gamma-ray sources such as Active Galactic Nuclei (AGNs) and transients like Gamma-Ray Bursts (GRBs). With these objectives, CoMET is designed to have a low energy threshold with a wide field-of-view of about \SI{2}{\steradian}, at a high altitude, and combines ALTO particle detectors with CLiC air-Cherenkov detectors. In this contribution, we focus on the ALTO particle detector array performance only. Water Cherenkov detectors are used for the detection of secondary particles in atmospheric air showers while scintillators serve as muon counters. A detailed study is presented through air-shower, detector and trigger simulations, followed by the reconstruction of the event parameters and the extraction of the signal (gamma-rays) from the background (cosmic-rays). We present the sensitivity of the ALTO detectors to a list of astrophysical sources using two SEMLA analysis configurations.
}

\FullConference{37$^{\rm{th}}$ International Cosmic Ray Conference (ICRC 2021)\\
		July 12th -- 23rd, 2021\\
		Online -- Berlin, Germany}

\begin{document}
\maketitle

\section{Introduction to the CoMET project}
\label{sec:intro}

The CoMET (Cosmic Multi-perspective Event Tracker) R\&D project aims to optimise the wide field-of-view ground-based technique for very-high-energy (VHE) gamma-ray observations, also see \cite{comet_icrc21}.
With the science goal of studying the extra-galactic sources, the CoMET detector array is being designed for the detection of \SI{200}{\GeV} to  \SI{50}{\TeV} gamma-rays. 
The ALTO detector array is part of the CoMET effort, and this proceeding only focuses on the expected performance of ALTO. 

When the VHE gamma-rays enter the Earth's atmosphere, they produce a cascade of secondary particles leading to the air shower development. The ALTO project aims to design a particle detector array to observe the secondary air-shower particles. 
These particle detectors will be placed at a high altitude ($\sim$\SI{5}{\km} a.s.l.) to reach a lower-energy threshold. 


\section{The ALTO particle detector array}
\label{sec:detector_design}

The ALTO particle detector array consists of 207 detector "clusters" placed in a circular area of 160~m diameter, see Figure \ref{fig:unit-array} (left panel). 
A cluster consists of 6 detector units (DU) and a unit consists of a water Cherenkov detector (WCD) with a liquid scintillator detector (SD) underneath, separated by a 25~cm concrete slab, see Figure \ref{fig:unit-array} (right panel).

The WCD is a \SI{2.5}{\m} high and \SI{4.15}{\m} wide hexagonal-shaped light-tight tank filled with $\sim$\SI{25}{\m^3} of water, which utilizes the water Cherenkov technique \cite{MILAGRO}. 
The SD is filled with Linear Alkyl Benzene (LAB) with small quantities of wavelength shifters (PPO, POPOP) \cite{LAB}. 
For the WCD, a super-bialkali 8" Hamamatsu PMT (R5912--100) is used while for the SD, we use a standard-bialkali 8" Hamamatsu PMT (R5912). 
A reflective crown is placed on top of the WCD PMT to increase the Cherenkov photon collection surface. 
The inner walls of the WCD are blackened to improve the accuracy of the particle arrival time from the PMT signal, thus helping the reconstruction of the primary particle direction.

\section{Simulation and the expected performance of the ALTO particle detectors}
\label{sec:simulation}

In order to study the expected performance of the proposed ALTO detector array, a detailed simulation study has been conducted. 
The atmospheric air-shower and detector simulation is introduced in the following Sub-section \ref{subsec:simu}. 
The analysis configuration and reconstruction of the shower parameters are summarized in Sub-section \ref{subsec:reco}. 
The analysis followed for event selection is explained in Sub-Section \ref{subsec:SEMLA}. 
The expected performance of the ALTO particle detectors and the spectral response for a list of simulated point-like gamma-ray sources at a zenith angle of $18^\circ$ are shown in Sub-section \ref{subsec:exp_perf}.

\subsection{The atmospheric air shower and detector simulation}
\label{subsec:simu}
\begin{figure*}[t]
    \centering
    \includegraphics[width=0.43\textwidth] {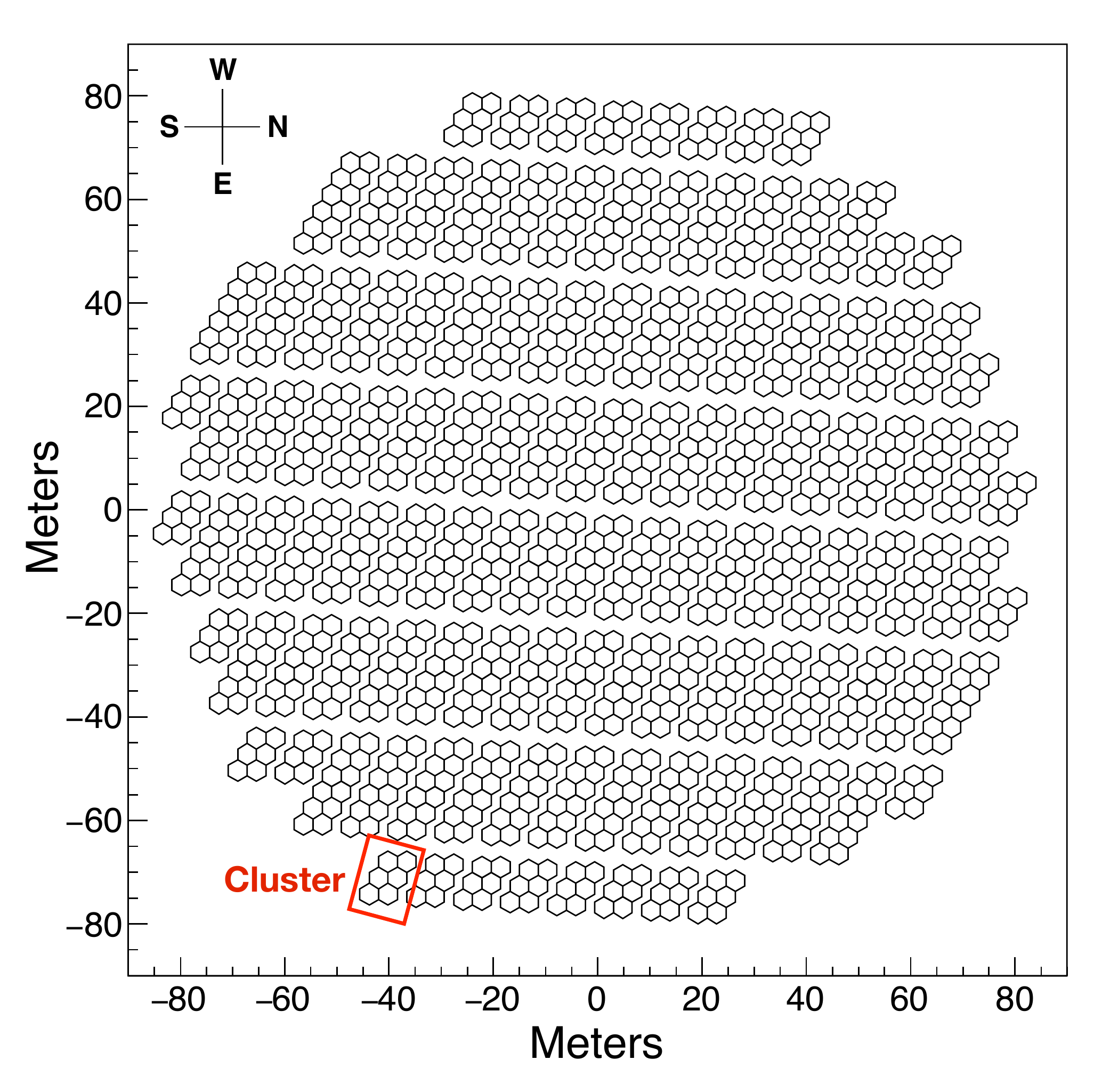}
    \hspace{0.1cm}
    \includegraphics[width=0.55\textwidth] {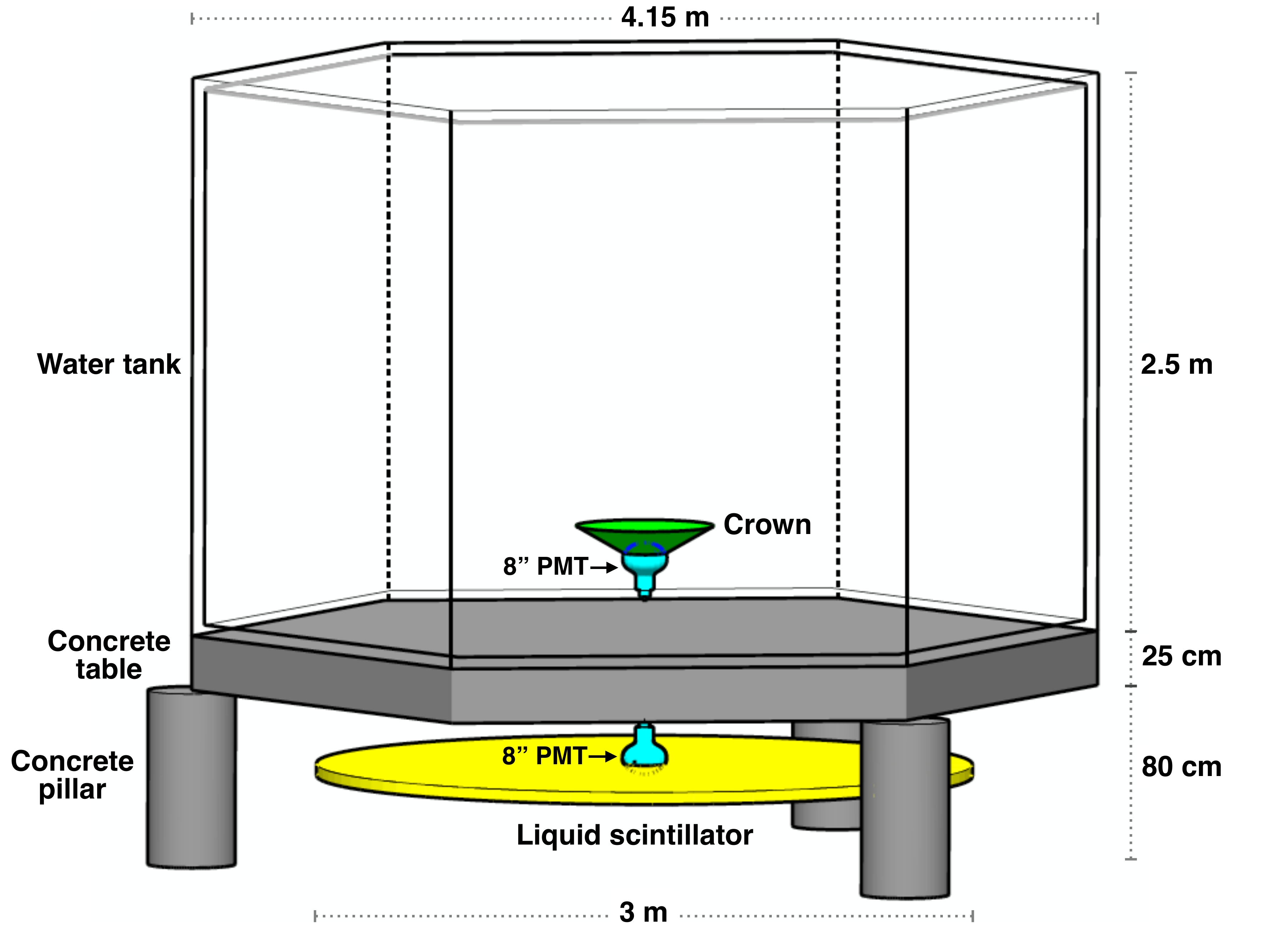}
    \vspace{-1 cm}
    \caption{\small{The proposed ALTO air shower particle detector array. \emph{Left panel:} The layout of the proposed array of $\rm 160\,m$ diameter. A \emph{cluster} consists of six ALTO detector units as highlighted in the red box. \emph{Right panel:} The geometry of a detector unit with its dimensions. This figure is obtained from \cite{SEMLA}. }}
    \label{fig:unit-array}
\end{figure*}

The atmospheric air shower development is simulated using the CORSIKA software (version 7.4387), while the GEANT4 libraries (version 4.10.02.p02) are used for computing detector response to the simulated showers. The parameters used in the air-shower simulation are summarized in Table \ref{table:simu}. 
For the air-shower simulation, the US standard atmosphere is used with the horizontal and vertical components of the magnetic field set as 21.12 $\mu$T and -8.25 $\mu$T, respectively. 
Figure \ref{fig:unit-array} shows the design of the proposed ALTO particle detector array used for the GEANT4 simulation.

\begin{table}[t]
    \centering
    \small
    \begin{tabular}{ccccccc}
        \hline
        \textbf{Primary} & \textbf{Power law} & \textbf{Energy} & \textbf{True} & \textbf{True} & \textbf{Impact} & \textbf{Events} \\ 
        
        \textbf{type} & \textbf{spectral} & \textbf{range} & \textbf{zenith [${}^{\circ}$] }& \textbf{azimuthal [${}^{\circ}$] } & \textbf{parameter} & \textbf{(x\boldmath{$10^6$)}} \\
        
        & \textbf{index} & \textbf{[TeV]} & \boldmath{$\theta_{\rm T}$} & \boldmath{$\phi_{\rm T}$} & \textbf{[m]} & \\ 
        
        \hline
            $\gamma$ ray & -2 & 0.01--100 & 18 & 0 & 0--130 & 34 \\
            proton & -2.7 & 0.06--100 & 15--21 & 0--360 & 0--184 & 224\\
        \hline
    \end{tabular}
    \vspace{-0.3 cm}
    \caption{\small{Parameters used in the CORSIKA air shower simulation}}
    \label{table:simu}
    
\end{table}

\subsection{Analysis configuration and reconstruction}
\label{subsec:reco}

We apply a "cluster trigger condition" to the simulated signals, where two-out-of-twelve detectors in a cluster should have a signal amplitude above 20~mV. 
After applying this trigger condition, we present and compare the following two analysis configurations: \emph{config-Q5} and \emph{config-Q2}, henceforth simply denoted as \emph{Q5} and \emph{Q2}, respectively. 
For \emph{Q5} the integrated charge of the WCD signal used in the reconstruction procedure and in the final analysis comprising the background rejection should be above 5 photo-electrons (p.e.), while for \emph{Q2} it should be above 2 p.e. 
All the results presented in \cite{SEMLA} correspond to \emph{Q5}. 
Here, we also explore the \emph{Q2} configuration, with the aim of understanding the effect of the charge threshold in the analysis. 

An event should trigger at least eight WCDs ($N_{\rm WCD} \geq 8$) in order to be considered for the reconstruction procedure, where we use the time-of-maximum and the integrated charge of the signal from the WCDs. 
The NKG lateral charge distribution function \cite{Kamata, Greisen} is used to reconstruct the shower core position, the size of the shower ($N_{\rm pe}$), and the Moli\`ere radius. The arrival direction is instead evaluated using the hyperbolic shower front model \cite{LOFAR, LOPES}. 

\subsection{Event selection and energy reconstruction}
\label{subsec:SEMLA}

In order to efficiently extract the $\gamma$-rays from the background of cosmic rays, we use the SEMLA (Signal Extraction using Machine Learning for ALTO) analysis strategy \cite{SEMLA} after the reconstruction procedure. 
The analysis involves successive filtering using artificial neural networks (Multi Layer Perceptron, MLP) implemented using the ROOT-TMVA package (ROOT version 6.08.06; TMVA version 4.2.0).
Based on the analysis configuration used (\emph{Q5} or \emph{Q2}), the same integrated charge threshold of 5 or 2 p.e. is also propagated to the SEMLA analysis.
The simulated data is split into two equal and unbiased halves, one of which is used for training/testing to obtain selection cuts and weights (\emph{ISet}), while the other independent half is used for studying the performance of the analysis (\emph{PSet}). 
The analysis consists of four successive stages (A, B, C and D) as introduced below.
\vspace{-0.3cm}
\begin{itemize}

     \item \textbf{Stage A:} The first step of the analysis is the event cleaning where some poorly-reconstructed events are removed. This is carried out by identifying abnormal values in the fit results and applying simple cuts.
     \vspace{-0.3cm}
     
     \item \textbf{Stage B:} This stage focuses on removing the poorly-reconstructed $\gamma$-rays using a classification technique. Here, a set of $\gamma$-ray events are tagged as well-reconstructed events (\emph{signal}), if their true impact parameter is within the ALTO array and the reconstructed shower direction and shower core are close to their true values. The rest of the gamma-ray events are tagged as poorly-reconstructed events (\emph{background}). With the classification training/testing process, the selection cuts for a given $\gamma$-ray efficiency of well-reconstructed events are obtained. 
     \vspace{-0.3cm}
     
     \item \textbf{Stage C:} The crucial part of the SEMLA analysis is to separate the $\gamma$-rays from the proton background through a classification procedure, which is performed in stage C. $\gamma$-ray events passing stage B are considered as the \emph{signal} and the proton events passing stage B are considered as the \emph{background}.  Similar to stage B, the selection cuts are obtained for a given $\gamma$-ray efficiency and applied to both the $\gamma$-rays and the protons.
     \vspace{-0.3cm}
     
     \item \textbf{Stage D:} In this stage, the primary particle energy for events passing stage C is reconstructed, using a regression model.

 \end{itemize}
  \vspace{-0.3cm}
After these four successive stages, the obtained selection cuts and weights are applied to the independent \emph{PSet}. 
The effects of each stage of the SEMLA analysis on the \emph{PSet} simulated events are shown in Table \ref{table:event_stat} for both the analysis configurations. 
While comparing the fraction of events between \emph{Q2} and \emph{Q5}, we notice that the SEMLA analysis effect is almost the same. 
However, we obtain more gamma-ray and proton events with \emph{Q2}, as indicated by the rates in Table \ref{table:event_stat}.

\begin{table}[t]
    \small
    \centering
    \begin{tabular}{c|cccc|cccc}
        \hline

        \textbf{Stages} & \multicolumn{2}{c}{\textbf{$\gamma$-ray fraction}}  & \multicolumn{2}{c}{\textbf{Proton fraction}} & \multicolumn{2}{c}{\textbf{$\gamma$-ray rate}} & \multicolumn{2}{c}{\textbf{Cosmic-ray rate}} \\
        
        &  & & & & \multicolumn{2}{c}{$\mathbf{[min^{-1}]}$} & \multicolumn{2}{c}{$\mathbf{[deg^{-2}min^{-1}]}$}\\
        
        \hline
        
        & \textbf{config-Q5} & \textbf{config-Q2} & \textbf{Q5} & \textbf{Q2} & \textbf{Q5} & \textbf{Q2} & \textbf{Q5} & \textbf{Q2}\\
        \hline
            Stage A & 95.2\% & 95.2\% & 88.9\% & 88.5\%& 2.11 & 2.43 & 171.0 & 213.6 \tabularnewline 
            Stage B & 36.6\% & 34.5\% & 25.8\% & 24\%& 0.81 & 0.88 & 49.2 & 57.8 \tabularnewline 
            Stages C--D & 25.1\% & 23.7\% & 1.7\% & 1.7\%& 0.56 & 0.61 & 3.3 & 4.1 \tabularnewline 
        \hline
    \end{tabular}
    
    \vspace{-0.3 cm}
    \caption{\small{The fraction of remaining simulated $\gamma$-rays and protons after each SEMLA stage relative for events with $N_{\rm WCD}$ $\geq$ 8. The $\gamma$-ray rates are estimated for the Pseudo-Crab source, see \cite{SEMLA}, while the cosmic-ray rates result from rescaling the protons to the cosmic-ray flux, as described in \cite{SEMLA}. The values presented here are obtained with the \emph{PSet} data sample.
    }}
    \label{table:event_stat}
    
\end{table}

\subsection{Expected performance of the ALTO particle detectors}
\label{subsec:exp_perf}

\begin{figure}[t]
    \centering
    
    \includegraphics[width=0.48\textwidth]{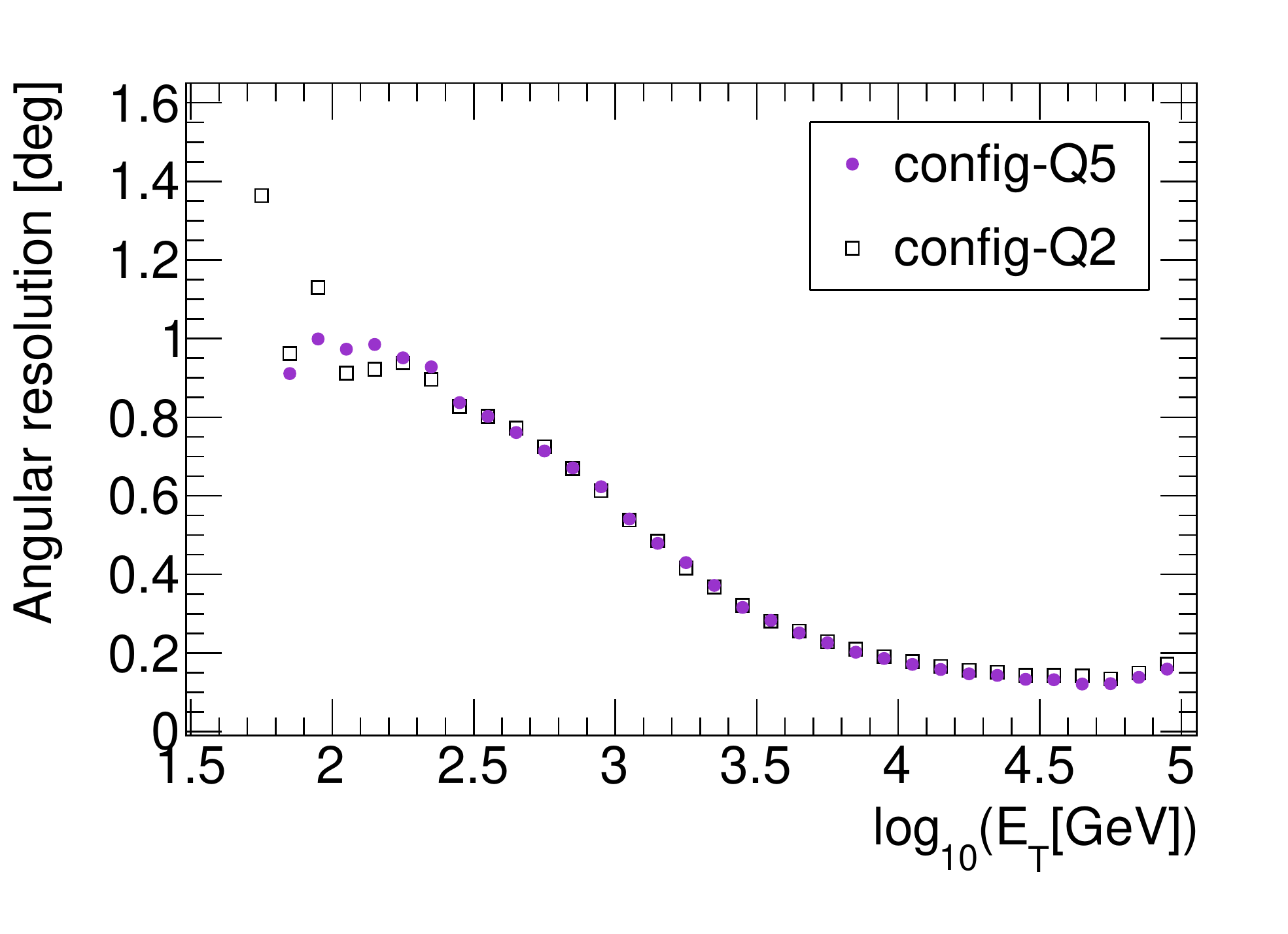}
     \hspace{.1 cm}
     \includegraphics[width=0.48\textwidth]{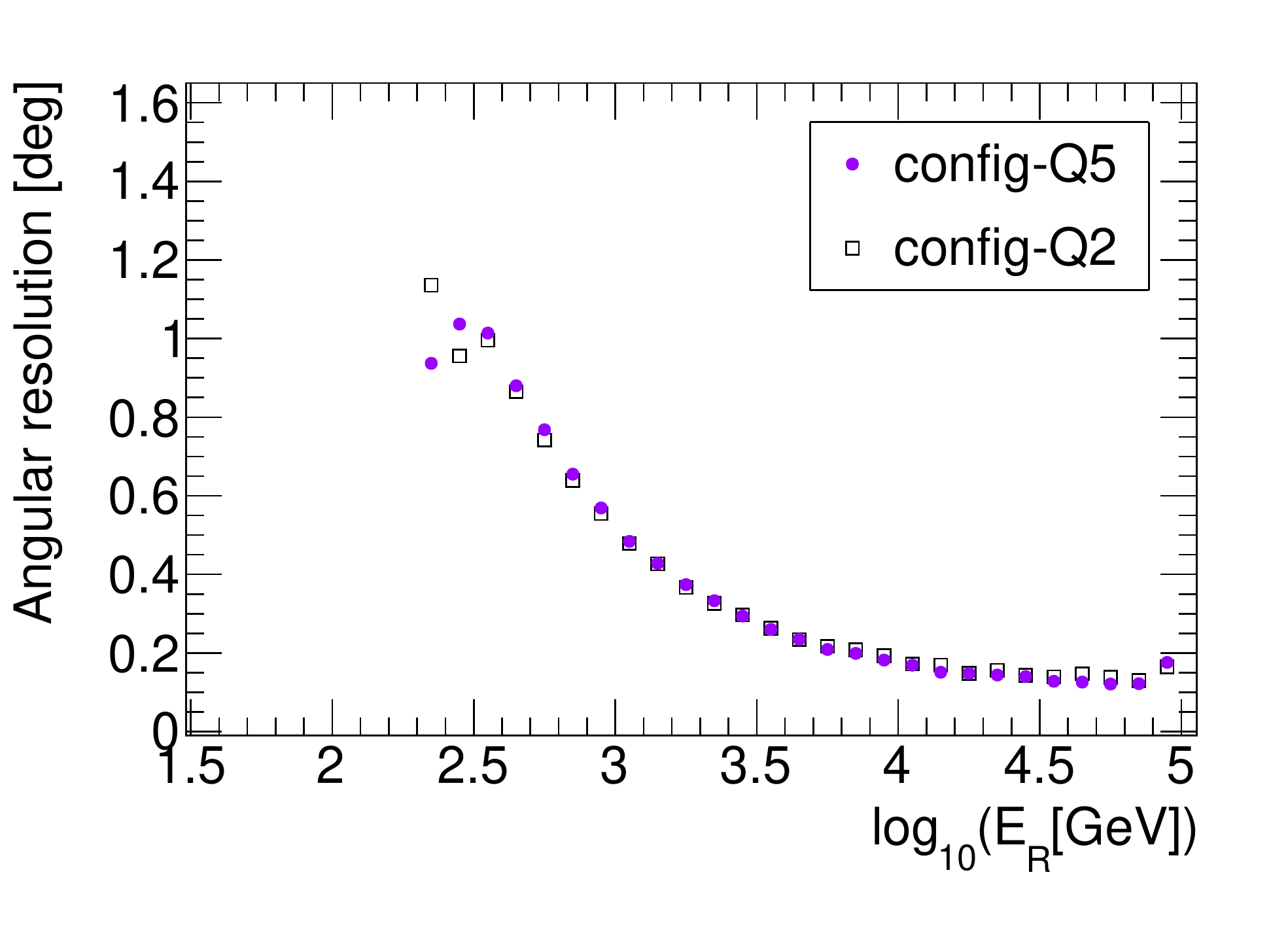}
     
     \vspace{-0.7 cm}
    \caption{\small {Angular resolution for a simulated point-like gamma-ray source at zenith 18$^{\circ}$ after the SEMLA analysis. The angular resolution is shown as a function of $\log_{10}(\rm E_{\rm T})$ (\emph{left panel}) and $\log_{10}(E_{\rm R})$ (\emph{right panel}).
    The plots are obtained using the \emph{PSet} data sample.}}

    \label{fig:ang_res}
    
\end{figure}

\begin{figure}[t]
    \centering
    
     \includegraphics[width=0.48\textwidth, trim={0 2.8cm 0 0}, clip]{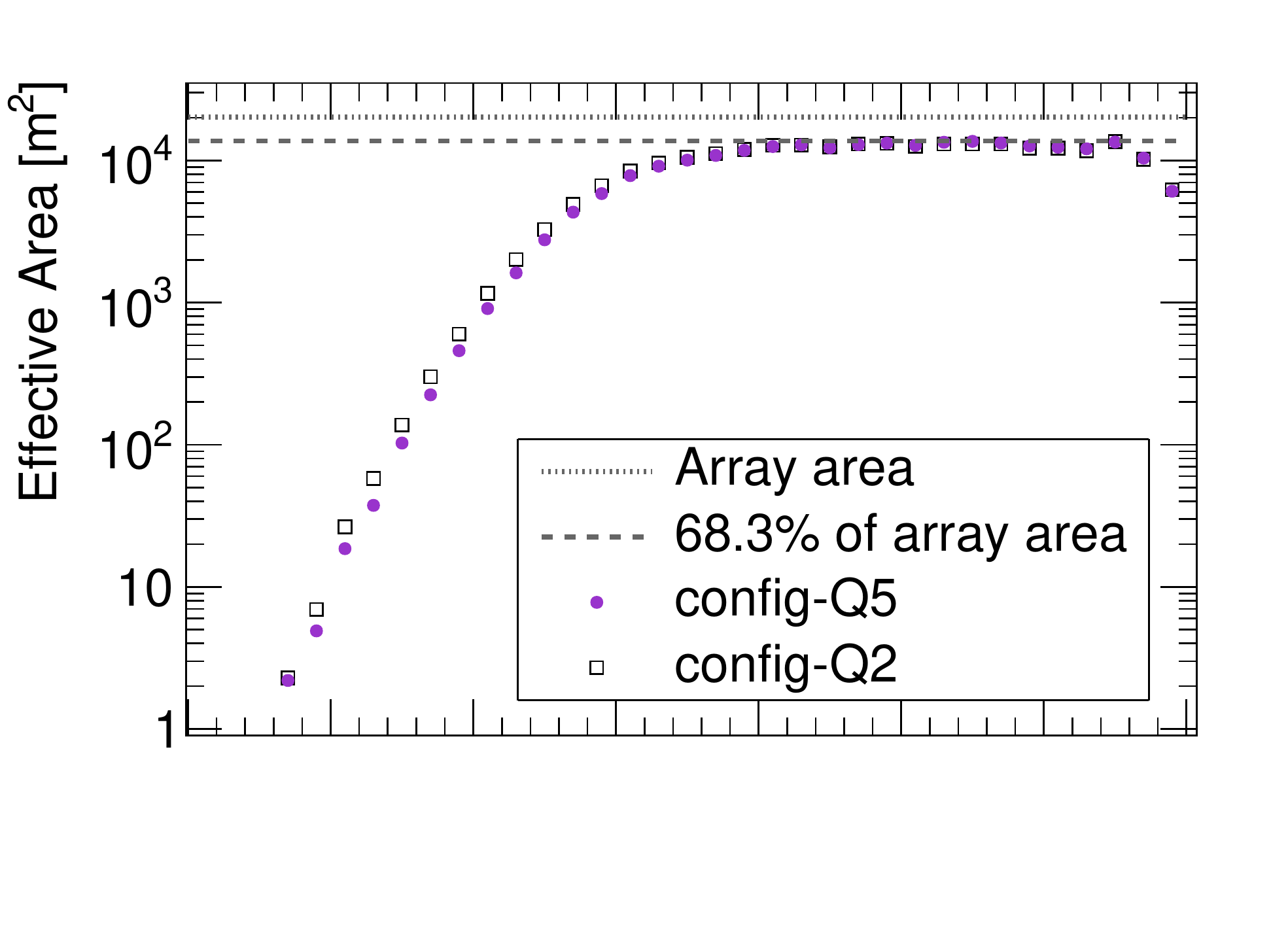}
     \hspace{.1 cm}
     \includegraphics[width=0.48\textwidth, trim={0 2.8cm 0 0}, clip]{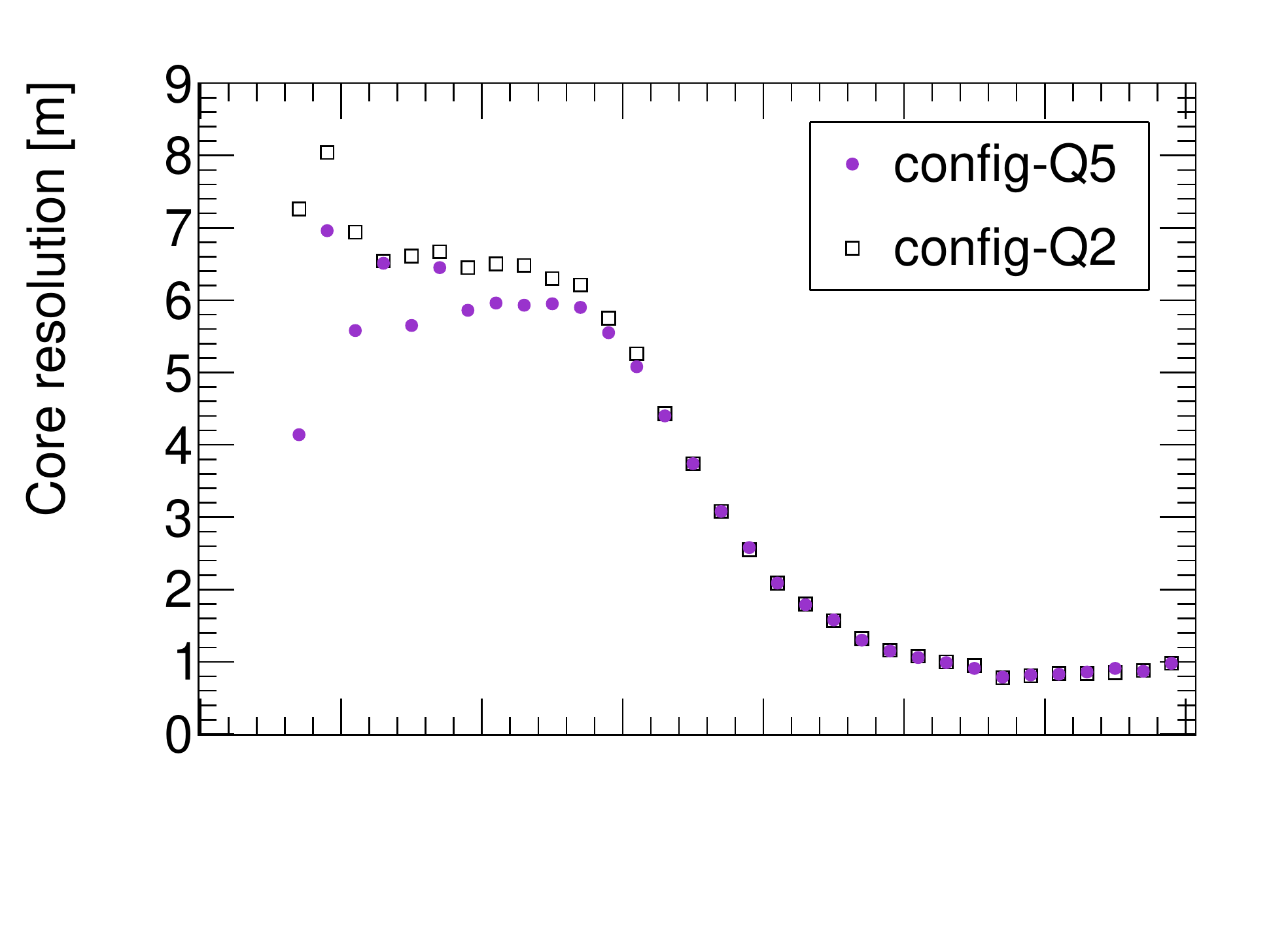}\\
     \vspace{-0.04cm} 
     
    \includegraphics[width=0.48\textwidth, trim={0 0 0 1cm}, clip]{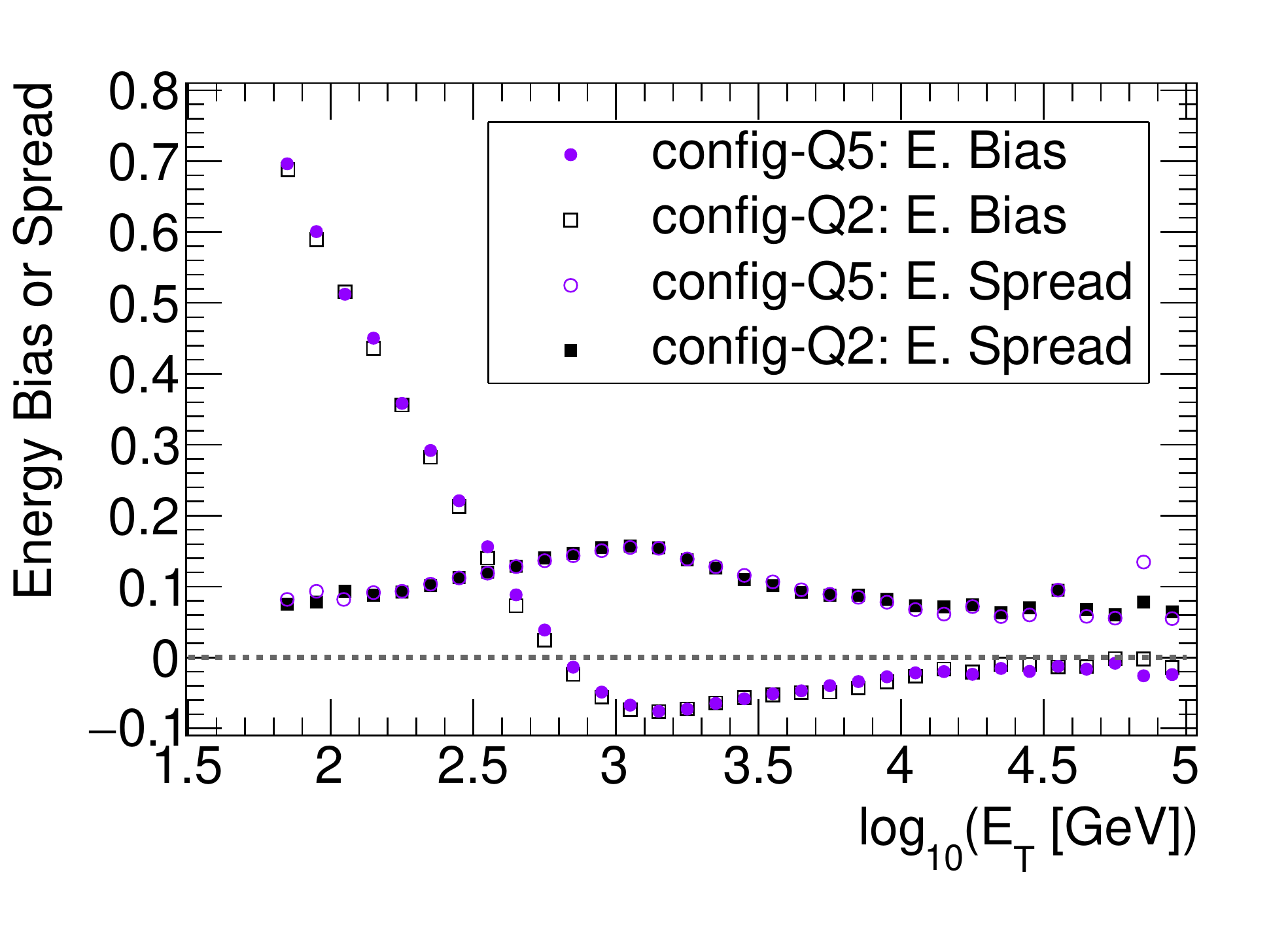}
     \hspace{.1 cm}
     \includegraphics[width=0.48\textwidth, trim={0 0 0 1cm}, clip]{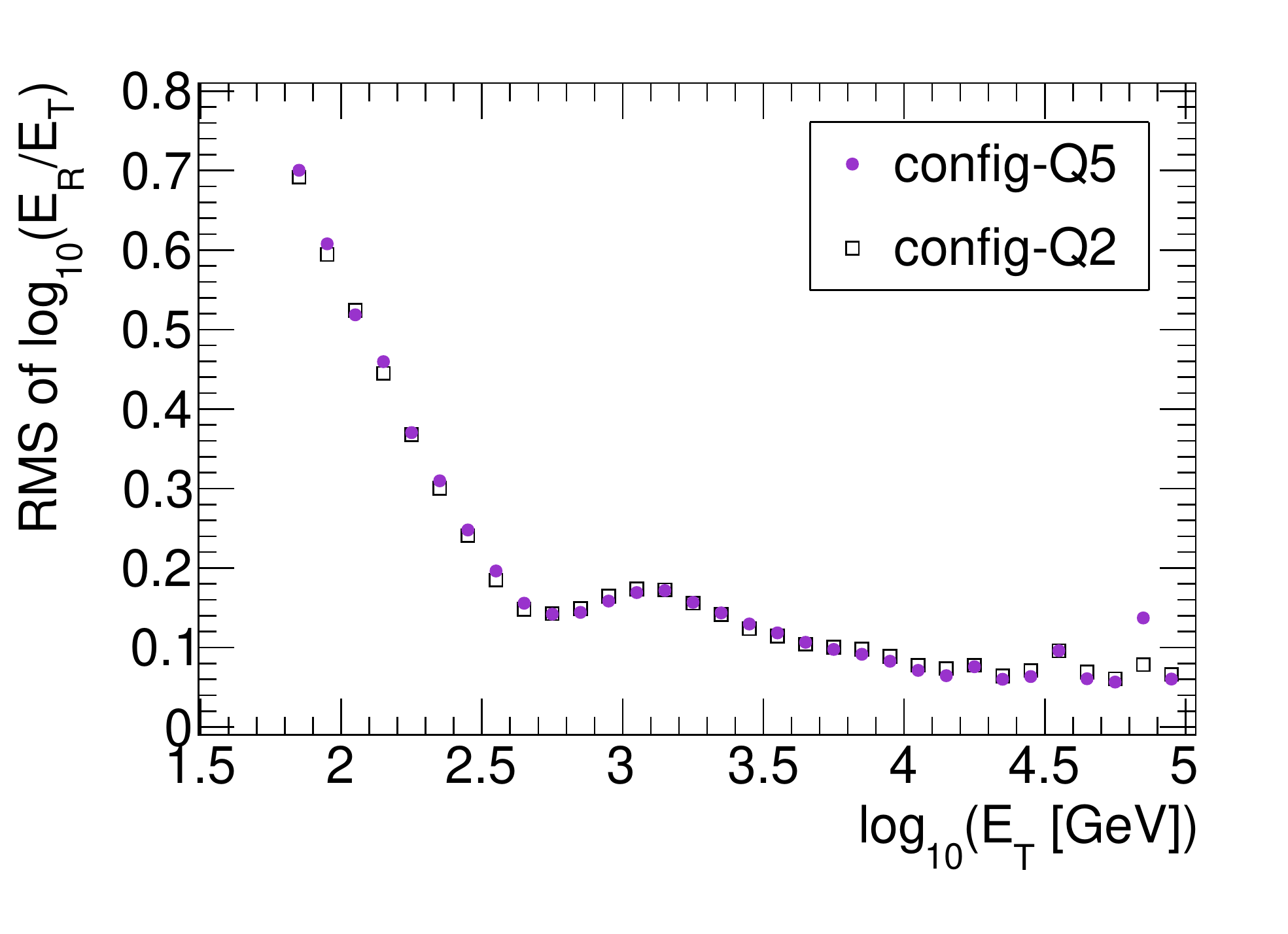}
     
     \vspace{-0.7cm}
    \caption{\small{The performance plots of ALTO for a simulated point-like $\gamma$-ray source at zenith 18$^{\circ}$ after the SEMLA analysis and the angular separation cut. The plots are obtained using \emph{PSet} data sample.}}
    
    \label{fig:IRFs}
    
\end{figure}

After selecting the golden events using the SEMLA analysis, the performance of the two configurations in terms of angular resolution \cite{SEMLA} is compared in Figure \ref{fig:ang_res}. 
We notice that for both the analysis configurations, the angular resolution as a function of true ($E_{\rm T}$) and reconstructed energy ($E_{\rm R}$) are comparable, resulting in $\sim$0.8$^{\circ}$ at 300~GeV and $\sim$0.15$^{\circ}$ at 20~TeV. 
The values in Figure \ref{fig:ang_res}, right panel, are used to define the angular separation cut (commonly known as the $\theta^2$-cut) to define the ON-source region for the sensitivity and the spectral response calculation. 
Following the representations used in \cite{SEMLA}, effective area, core resolution, energy bias, spread and RMS are obtained, and they are compared for both analysis configurations in Figure \ref{fig:IRFs}. 
While comparing the effective areas of \emph{Q5} and \emph{Q2}, we find a slight gain with \emph{Q2} at the lowest energies. 
The energy bias, spread and RMS for both configurations look similar at all the energies, while the core resolution is slightly better for \emph{Q5} at the lowest energies.

\begin{figure}[t]
    \centering
    
    \includegraphics[width=\textwidth]{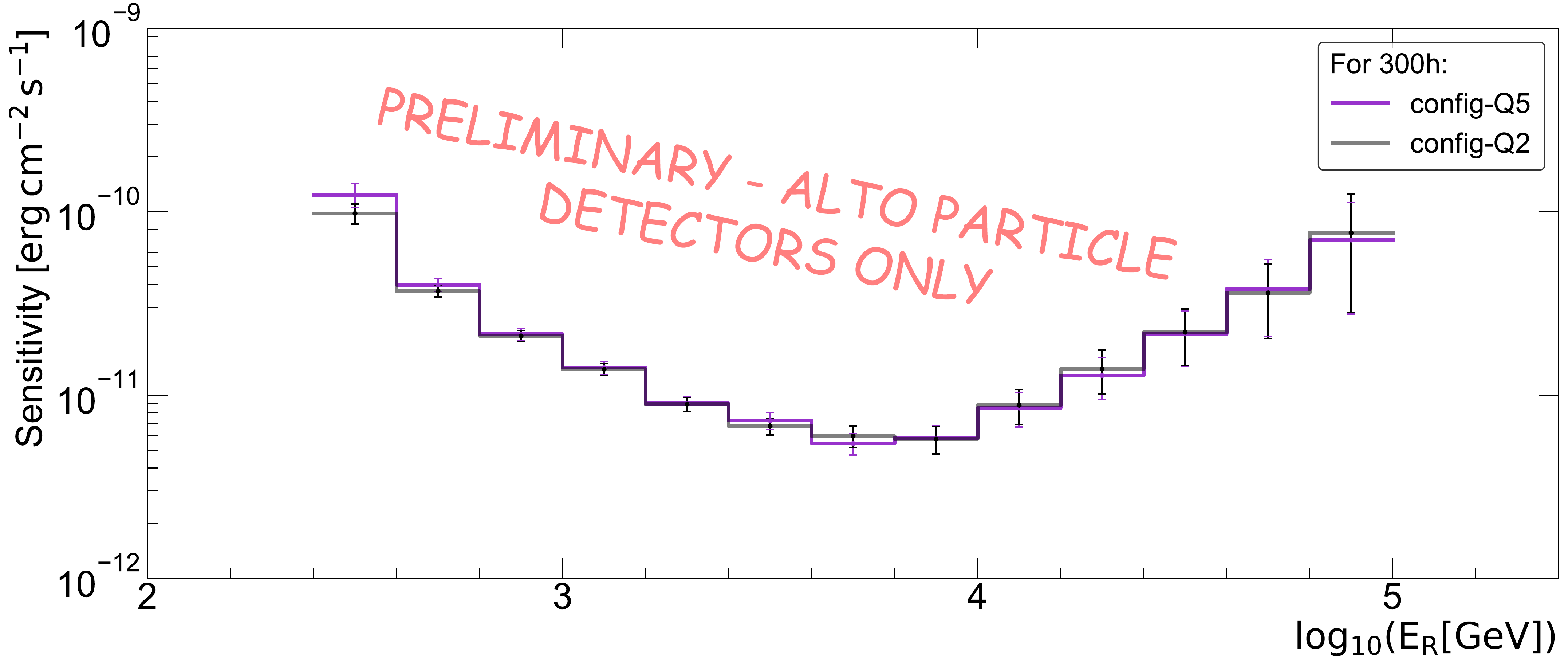}
     
    \caption{Expected ALTO detector array sensitivity for a point-like gamma-ray source described by a power-law spectral index of 2, at zenith 18$^{\circ}$ for 300 hours.The sensitivity plot is preliminary, because only the proton component of the cosmic-ray spectrum is considered here.}
    
    \label{fig:compare_sensitivity}
    
\end{figure}

The OFF-source regions for the background estimation are defined in 12 circular regions with a radius of $3^{\circ}$ at the same zenith angle. 
For the calculation of the sensitivity, we first find the rate of background events in the OFF regions  in each $\log_{10}(E_{\rm R})$ bin, for the simulated livetime.
Then, we find the rate of gamma-rays 
in the ON region, in each of the same $\log_{10}(E_{\rm R})$ bins.
Finally, for a given source spectrum we find the observation time required  
to reach a significance of $5\sigma$ (based on eqn.\ 17 in \cite{Li-Ma}), 
where $\alpha$ in that equation is the ratio of the area of the ON region to that of the OFF regions, 
and $N_{\rm ON}$, $N_{\rm OFF}$ are calculated from the respective rates and areas.
The expected ALTO particle detector array sensitivity for a point-like gamma-ray source described by a power-law spectral index of 2, at a zenith angle of 18$^{\circ}$, for 300 hours, is shown in Figure \ref{fig:compare_sensitivity} for both the configurations.

For the spectral response estimation, the rate of gamma rays and background in each $\log_{10}(E_{\rm R})$ bin is obtained for the given input spectrum.
Then following a similar procedure to the point-source sensitivity estimation, the significance (based on eqn. 5 in \cite{Li-Ma}) of the source is estimated. The time taken for various sources to reach an integrated significance of $5\sigma$ is shown in Table \ref{table:spectral_response}. The expected spectral response shows that the ALTO particle detector array would be able to detect the soft spectrum gamma-ray sources in a reasonable time-scale.

\begin{table}[h]
    \centering
    \begin{tabular}{*6c}
        \hline

        \textbf{Source} & \multicolumn{4}{c}{\textbf{Model}}  & \textbf{Approximate time to} \\
        
        &  &  &  &  &  \textbf{reach a 5-$\sigma$ detection}\\

        \hline
        &  \multicolumn{4}{c}{\textbf{Power law}} & \\
        \hline 
        
        &  & \bm{$\Gamma$} & \bm{$N_{0}$}\, \bm{$\times10^{-12}$} & \bm{$E_{0}$} & \\
        
        GRB~180720B* \cite{grb1} & & 3.6 & 5 $\times10^5$ &  0.23 & $\sim$ 38~seconds \tabularnewline
        
        PKS~2155--304 flare \cite{2010_pks2155, 2017_pks2155} &  & 3.2 & 191 & 1 & $\sim$ 24~minutes \tabularnewline
        
        GRB~190114C \cite{grb2} & & 5.4 & 2500 &  0.35 & $\sim$ 31~minutes \tabularnewline
        
        PG~1553+113 flare \cite{pg1553} & & 4.9 & 0.54 & 1 & $\sim$ 21~days \tabularnewline
        
        \hline
        & \multicolumn{4}{c}{\textbf{Log-parabola}} & \\
        \hline
        
        &  \textbf{A} & \textbf{B} & \bm{$N_{0}$}\, \bm{$\times10^{-12}$} & \bm{$E_{0}$} & \\
       
       Crab Nebula \cite{crab} & 2.1 & 0.24 & 179 & 0.521 & $\sim$ 17~hours \tabularnewline
       
       PKS 2155-304 quiescent \cite{2017_pks2155} & 3.21 & 0.16 & 4.11 & 1 & $\sim$ 33~days \tabularnewline 
      
        \hline
        
    \end{tabular}
    
    \caption{The expected spectral response for various VHE point-like gamma-ray sources at zenith 18$^{\circ}$, obtained using \emph{config-Q2}. For the power law model, $\Gamma$, $N_{0}$ and $E_{0}$ represent  the spectral index, the normalisation factor ($ph\,cm^{-2}\,s^{-1}\,TeV^{-1}$) and the energy scale factor ($TeV$) respectively. The labels A and B are the log-parabola parameters. For GRB~180720B, $N_{0}$ is obtained by extrapolating its light curve to the time of alert.}
    
    \label{table:spectral_response}
    
\end{table}

\section{The ALTO prototype at Linnaeus University}
\label{sec:prototype}

A prototype made of 2 DUs has been built on the campus of Linnaeus University, V\"axj\"o, Sweden (160m a.s.l.). Each DU consists of a water tank and a scintillator tank underneath a 25cm-thick concrete slab; their centres are separated by \SI{6}{\m}. Several independent small scintillation detectors have been added in order to monitor the stability of the prototype response using atmospheric muons. 
From February 2019 to September 2020, the ALTO DUs and the accompanying monitoring detectors have been continuously operated in
a trigger condition similar to the one described in this proceeding, {\it i.e.,} two detectors should have a signal amplitude above 20 mV to store data. 

Over the period, the ambient temperature ranged from $-20^\circ$C to $32^\circ$C, while the water temperature stayed within  6--$23^\circ$C and the scintillator one within 0--$30^\circ$C (for a flash point of $140^\circ$C). Heating wires immersed in the water were able to provide \SI{180}{\watt} per water tank during Winter.

The single rate (counting rate of a detector beyond 20~mV), was 1.2~kHz for water tanks and 20--\SI{26}{\kilo\hertz} for scintillator tanks. The high value of the latter originates from ground natural radioactivity. Indeed, when the scintillator tanks were moved on the top of the concrete slab, for the prototype operation in a different configuration, their single rate fell down to 3.3--4.7~kHz.

An ``event'' is defined when at least 2 PMTs exceed the threshold of 20~mV within a time window of 120~ns. The event rate of the prototype was 850~Hz, made of $\approx 350$~Hz of \emph{muon} rate from each detection unit, and $\approx 4$~Hz of \emph{shower} rate.  A \emph{muon} is defined as an in-time signal in both water and scintillator tanks of a DU, with no signal in the other DU, and a \emph{shower} is defined as a coincident signal in both water tanks. The remaining rate was due to random coincidences from scintillator tanks.

In its different configurations, the prototype is in operation since May 2018, {\it i.e.,} for 3 years. 
It is used for several measurements such as stability and ageing of the elements and mediums: the PMT and HV-generating socket housed in PMMA containers in water tanks, the reflective crown, the water and  liquid scintillator, the tank structure, and the data acquisition device \cite{WAVECATCHER}. The prototype also allows to test our calibration procedures to measure the PMT gain and the time offsets on standard data. Since October 2020, the prototype also includes air-Cherenkov detectors for CoMET \cite{comet_icrc21} tests and characterization.

\section{Conclusion}
In this contribution, we demonstrate that the ALTO particle detector array would be able to detect extra-galactic VHE gamma-ray sources such as active galactic nuclei and GRBs in minute to hours time-scales. This is achieved by the current hardware design, proposed altitude, reconstruction, and analysis method used. The existing performance can be enhanced by coupling the ALTO particle detectors with the Cherenkov Light Collectors (CLiC) during darkness, presented in \cite{comet_icrc21}.

\section{Acknowledgements}

\url{https://alto-gamma-ray-observatory.org/acknowledgements/}



\end{document}